\documentclass[aps,floats,prd,twocolumn,nofootinbib,superscriptaddress]{revtex4-2}

\usepackage{amssymb}
\usepackage{amsmath}

\usepackage{graphicx}
\usepackage{graphics}
\usepackage{dcolumn}
\usepackage{color}
\usepackage{fancyhdr} 
\usepackage{graphicx}

\usepackage{subfig}

\usepackage[utf8]{inputenc}

\usepackage[justification=centerlast, format=plain, labelfont=bf]{caption}

\def\VEV#1{\left\langle #1 \right\rangle}

\newcommand{\be}{\begin{equation}}
	\newcommand{\ee}{\end{equation}}
\newcommand{\ba}{\begin{align}}
	\newcommand{\ea}{\end{align}}

\newcommand{\Mpcinv}{ {\rm Mpc}^{-1} }

\begin{document}
	
	\title{Cosmic Variance of the 21-cm Global Signal}
	
	\author{Julian B.~Mu\~noz}
	\email{julianmunoz@fas.harvard.edu}
	\affiliation{Department of Physics, Harvard University, 17 Oxford St., Cambridge, MA 02138, USA}
	\author{Francis-Yan Cyr-Racine}
	\email{fycr@unm.edu}
	\affiliation{Department of Physics and Astronomy, University of New Mexico, 210 Yale Blvd NE, Albuquerque, NM 87106, USA}

	\date{\today}
	
	\begin{abstract}
		Cosmological measurements of the 21-cm line of neutral hydrogen are poised to dramatically enhance our understanding of the early universe.
		In particular, both the epochs of reionization and cosmic dawn remain largely uncharted, and the 21-cm signal is one of the few 
		probes to reach them.
		The simplest 21-cm measurement is the global signal (GS), which corresponds to the averaged absorption or emission of 21-cm photons across the entire sky.
		While bright radio foregrounds swamp the cosmic signal over the entire frequency range observable, presenting a formidable hurdle,
		they can in principle be subtracted, given enough sensitivity.
		Here, however, we point out an additional---and irreducible---source of uncertainty for the 21-cm GS: cosmic variance.
		The cosmic-variance noise arises from the finite volume of the universe accessible to 21-cm experiments.
		Due to the cosmological redshifting of 21-cm photons, each observed frequency probes our universe during a particular cosmic age, corresponding to a narrow redshift slice.
		The presence of large 21-cm fluctuations 
		makes the GS within each slice different than the GS averaged over the entire universe.
		We estimate the size of this cosmic-variance noise, and find that for a standard scenario it has a size of $\sim 0.1$ mK, which is $\sim 10\%$ of the size of the expected instrumental noise of a year-long experiment.
		Interestingly, cosmic variance can overtake instrumental noise for scenarios with extreme 21-cm fluctuations, 
		such as those suggested to explain the sharpness of the claimed EDGES detection.
		Moreover, as large-scale 21-cm fluctuations are coherent over long distances, cosmic variance correlates the measurements of the GS at nearby redshifts, leading to off-diagonal uncertainties that have so far been neglected.
	\end{abstract}

	\maketitle

	The first stars formed a few hundred million years after the big bang, during the epoch we call cosmic dawn. 
	Their birth sourced abundant Lyman-$\alpha$ radiation, which allowed hydrogen to absorb 21-cm photons from the cosmic-microwave background (CMB).
	Subsequently, X-rays heated up the intergalactic hydrogen, which prompted it to emit 21-cm photons, while ultraviolet photons progressively ionized it until no hyperfine transitions were possible.
	Tracing the evolution of this 21-cm signal across cosmic dawn, which roughly covers the redshift range $z\approx 12-25$ ($100-400$ million years after the big bang), and the succesive epoch of reionization (EoR), at $z\approx 6-12$ (up to a billion years after the big bang), is imperative to understand of the astrophysics of the early universe~\cite{Madau:1996cs,Barkana:2000fd,Yoshida:2003rw,Furlanetto:2006jb,Pritchard:2008da,Pritchard:2011xb,loeb2013first,Bromm:2011cw}.

	The most straightforward 21-cm measurement is the so-called global signal (GS),
	which traces the average absorption or emission of 21-cm photons across the entire cosmos~\cite{Shaver:1999gb,Furlanetto:2006tf,Pritchard:2010pa,Mirocha:2015jra,Cohen:2016jbh,Liu:2015gaa}.
	A broad landscape of experiments are targeting this signal, such as EDGES~\cite{Bowman:2018yin}, LEDA~\cite{LEDA}, SARAS~\cite{Singh:2017syr}, PRIzM~\cite{PRIZM}, and SCHI-HI~\cite{Voytek:2013nua}.
	Their main obstacles are radio foregrounds (mainly Galactic synchrotron emission), which shine brighter than the cosmic 21-cm signal in the radio band~\cite{Rogers:2008vh,Bernardi:2009pi,deOliveiraCosta:2008pb,Bernardi:2016pva}.
	Thus, any analysis ought to simultaneously subtract these large foregrounds from the data when searching for the cosmological signal.
	Moreover, the presence of bright foregrounds, even if adequately subtracted, leaves thermal (Poissonian) noise in the cleaned data~\cite{Harker:2011et,Liu:2012xy}.
	This noise can be reduced by increasing the observation time, which allows for a cosmological detection of the 21-cm GS.

\begin{figure*}[hbtp!]
		\includegraphics[width=0.8\textwidth]{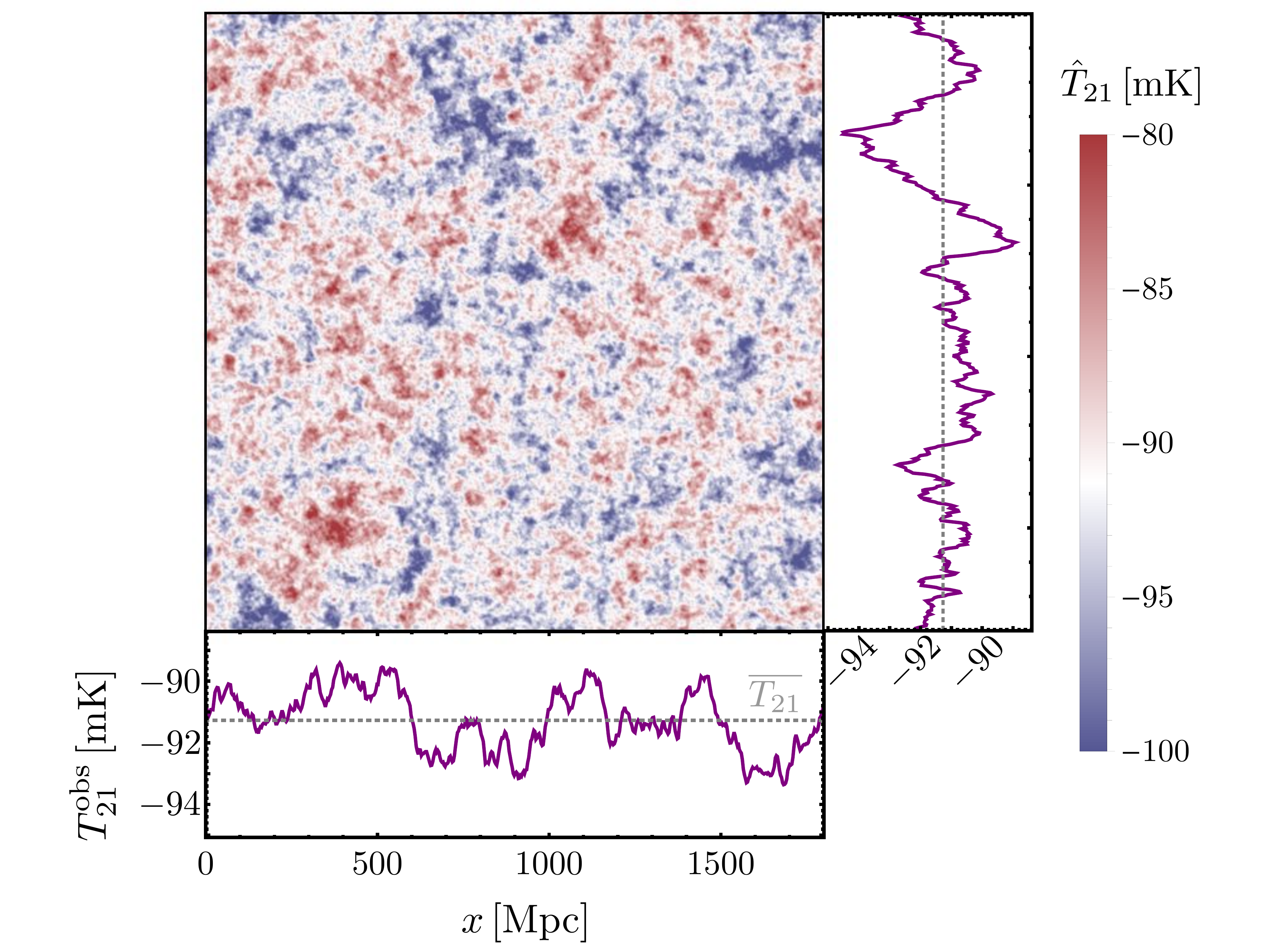}
		\caption{
			The heat map shows the 21-cm temperature averaged over one of the directions of our simulation, $\hat T_{21}$, at redshift $z=16.8$.
			This map is further averaged over one more of the directions to obtain the purple lines in the side panels, which correspond to the 21-cm global signal (GS) that would be observed by an experiment with a 0.1 MHz bandwidth (which yields slices 3 comoving Mpc in width).
			The gray dotted line shows the ``true" GS, $\overline{T_{21}}$, obtained by averaging over the entire box.
			This figure illustrates how the GS measured over a thin slice can significantly depart from the true GS, giving rise to cosmic variance.
			Moreover, the GS measured at nearby slices (corresponding to close-by redshifts) is correlated, for distances as high as $\sim 100$ Mpc.
		}	
		\label{fig:illustration}
	\end{figure*}

	In this {\it Letter} we will show that, in addition to thermal noise, the 21-cm GS suffers from cosmic variance, which produces an irreducible---and previously neglected---source of noise.
	A 21-cm GS experiment does not have access to the entire volume of the universe at each observed frequency $\nu$, as the universe evolves over time.
	At a particular $\nu$ (or equivalently, redshift $z$) only a small slice of the universe is integrated to obtain the 21-cm GS.
	The particular value measured is, thus, drawn from a random distribution around the true GS, albeit with a non-zero variance due to the 21-cm fluctuations.
	This is illustrated in Fig.~\ref{fig:illustration}, where we show the output of a 21-cm simulation averaged over two of the three physical dimensions, which reproduces the procedure of measuring the GS.
	We only have access to one of such measurements, which need not coincide with the true GS, as they fluctuate around it, at the percent-level for this simulation.
	This is akin to other cosmological observables (such as galaxy or cluster counts~\cite{Somerville:2003bq,Trenti:2007dh,Moster:2010hf,Codis:2016dyz}, their correlation functions~\cite{Krause:2016jvl}, weak-lensing maps~\cite{Gruen:2015xxa}, and more famously the CMB~\cite{White:1993jr}), where the finite cosmic volume observed presents a noise floor, which however had not been computed before for the 21-cm GS (although it had for the 21-cm power spectrum~\cite{Pober:2013jna,Shaw:2019qin}).
	Moreover, as clear from Fig.~\ref{fig:illustration}, the 21-cm GS that would be measured at adjacent distances is correlated, as the same long-wavelength modes affect them, giving rise to cosmic covariance between measurements of the 21-cm GS at nearby redshifts.

	\newpage

	We begin defining the relevant quantities that we will use throughout this work, and how we will calculate them.
	Our observable is the 21-cm brightness temperature $T_{21}$, given by the amount of photons that neutral hydrogen absorbs from the CMB, if $T_{21}<0$, or emits, if $T_{21}>0$.
	Throughout this text we will obtain this quantity from {\tt 21cmvFAST} quasi-numerical simulations~\cite{Munoz:2019rhi,Munoz:2019fkt},
	based on {\tt 21cmFAST}~\cite{Mesinger:2007pd,Mesinger:2010ne,Greig:2015qca,Greig:2017jdj},
	which we discuss in more detail in Appendix~\ref{app:simsspecs}, including our choice of fiducial parameters.

	The 21-cm GS $\overline{T_{21}}(z)$ is defined as the average 21-cm temperature across the universe at each redshift $z$.
	Therefore, the 21-cm temperature at any point $\mathbf x$ can be generically decomposed as
	\be
	T_{21}(\mathbf x,z) = \overline{T_{21}}(z) + \delta T_{21}(\mathbf x,z),
	\label{eq:T21xdef}
	\ee
	where $\delta T_{21}(\mathbf x,z)$ is the 21-cm fluctuation.
	In practice, however, we do not have access to the entire universe at each $z$.
	Points further from us are observed at earlier cosmic times, and thus at higher $z$.
	Measuring the GS at a particular $z$ then implies integrating over a thin shell of the universe at a comoving distance $\chi(z)$ away from us.
	Mathematically, the observed 21-cm GS is given by
	\be
	T_{21}^{\rm obs}(z) = \int d^3 \mathbf x \, W_z(\mathbf x) T_{21}(\mathbf x, z),
	\label{eq:T21obs}
	\ee
	where $W_z(\mathbf x)$ is the window function, which accounts for the geometry of the finite observation region.
	A simple example, and the one on which we will focus, is that of a 21-cm experiment observing the full sky, with a top-hat  selection function in the radial direction with width $\Delta \chi \ll \chi(z)$, although our formalism holds for any selection function.

	Integrating over only part of the universe at each $z$ means that the 21-cm fluctuation $\delta T_{21}$ need not average out, which contaminates our GS measurement.
	We illustrate this point in Fig.~\ref{fig:illustration}, where we show the 21-cm signal from one of our simulations averaged over thin slices across either the $x$ or $y$ directions.
	Each of these slices provides an estimator for the 21-cm GS, $T_{21}^{\rm obs}$, which clearly varies from one slice to another, illustrating how having access to a finite cosmological volume, and a single realization of the universe, produces an intrinsic variance to the GS.
	This is an example of cosmic variance.

	We find the size of the cosmic variance by studying how much $T_{21}^{\rm obs}$ fluctuates around the true GS.
	First, it is clear from Eqs.~(\ref{eq:T21xdef},\ref{eq:T21obs}) that the ensemble average (denoted by brackets) of the observed global signal is unbiased, $\VEV{T_{21}^{\rm obs} } = \overline{T_{21}}$, by construction.
	There will be, however, a nonzero variance for our estimator $T_{21}^{\rm obs}$.
	This variance is given by the autocorrelation (i.e., the zero-lag two-point function) of the observed 21-cm GS,
	\be
	\sigma_{21}^2 (z) = \VEV{ [T_{21}^{\rm obs}(z)]^2 } - \VEV{T_{21}^{\rm obs} (z) }^2,
	\label{eq:var1}
	\ee
	which can be computed in Fourier space as
	\be
	\sigma_{21}^2(z) = \int \dfrac{d^3k}{(2\pi)^3} P_{21}(k,z) \mathcal W_z^2(\mathbf k),
	\label{eq:sigma21general}
	\ee
	in terms of the power spectrum $P_{21}$ of the 21-cm fluctuations. 
	Here, $\mathcal W_z(\mathbf k)$ is the Fourier transform of the window function $W_z(\mathbf x)$, which for our simple radial top-hat is given by $\mathcal W_z(k) \approx j_0[k \chi(z)]$, as we demonstrate in Appendix~\ref{app:windows}.
	This result is isotropic in $k$, as we are integrating over the entire sphere, although the same is true for half of the sphere, which is closer to the actual selection function of a global-signal experiment.

	Eq.~\eqref{eq:sigma21general} is the key result of this work, and it encapsulates the main insight: the 21-cm fluctuations produce an irreducible source of theoretical noise on the global signal.
	In order to evaluate this cosmic-variance noise we ought to know the 21-cm power spectrum $P_{21}$, which we obtain through {\tt 21cmvFAST} simulations.
	In particular, large-scale fluctuations (with small $k$) are most important, as small-scale (large-$k$) modes are averaged within the observed region, so the window function in Eq.~\eqref{eq:sigma21general} suppresses their contribution to the integral.
	Large-scale modes are difficult to measure in simulations, due to finite-volume effects.
	In order to model them, we will approximate the 21-cm fluctuations as tracing the matter overdensities $\delta_m$ at large scales
	\be
	\delta T_{21} (\mathbf x,z) = b_m (z) \delta_m(\mathbf x,z),
	\label{eq:matterbias}
	\ee
	with a bias coefficient $b_m(z)$ that we fit to our simulation results.
	As our simulations have low noise for large $k$, we divide our data---and thus all integrals---into two regimes: for $k\geq0.02\,\Mpcinv$ we will directly interpolate from our simulations, whereas for $k<0.02\,\Mpcinv$ we will fit for $b_m$ to overcome the simulation noise, although we have checked that interpolating between simulation data-points returns the same integral within 10\%.
	We detail this procedure in Appendix~\ref{app:largescale}, where we also show the validity of Eq.~\eqref{eq:matterbias}.

\begin{figure}[hbtp!]
		\centering
		\includegraphics[width=0.48\textwidth]{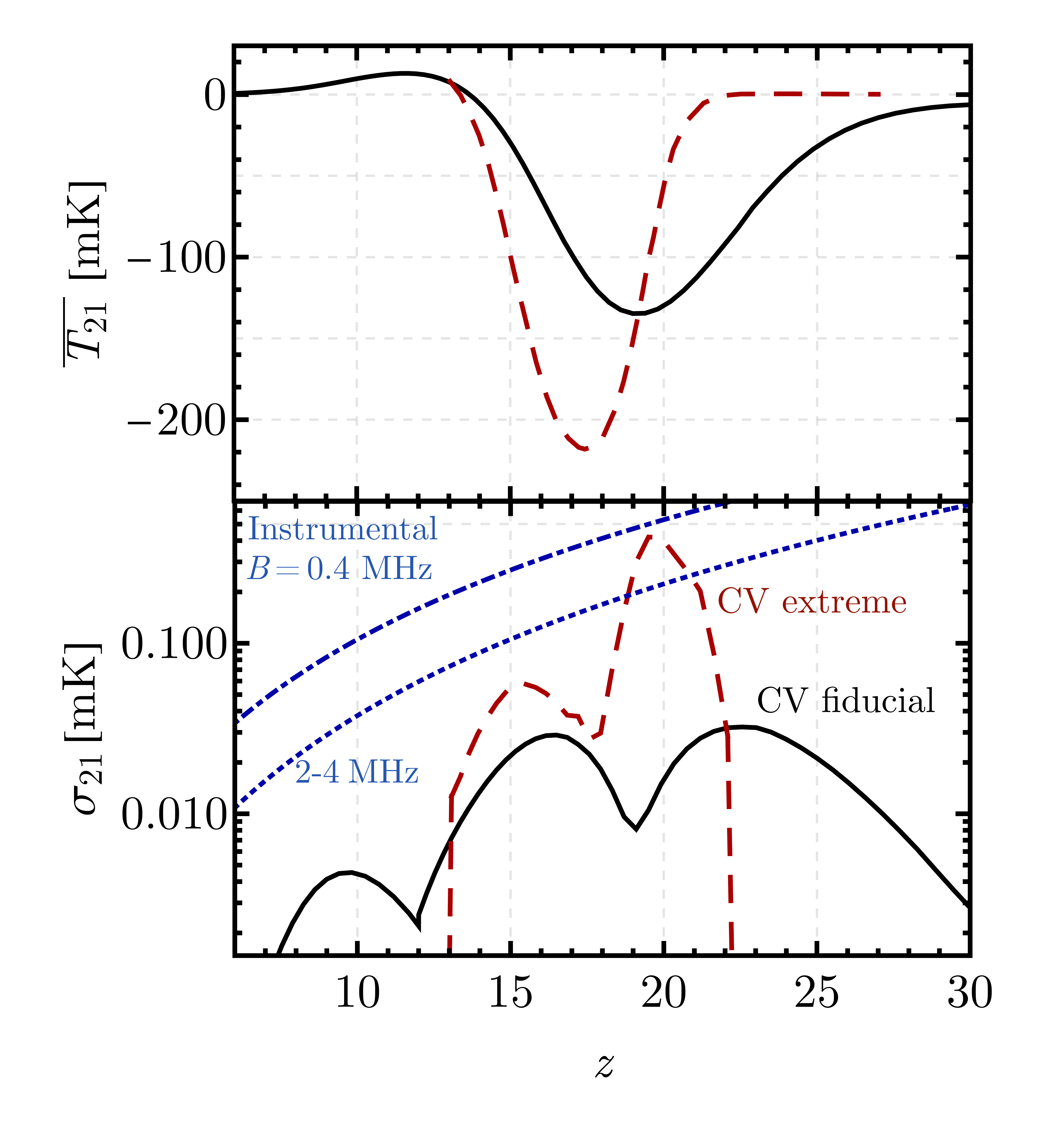}
		\caption{{\bf Top:} Global signal as a function of redshift $z$ for our fiducial model (in black) and that of Ref.~\cite{Kaurov:2018kez} (in red dashed), which was designed to grow extremely fast to fit the sharpness of the EDGES detection.
			{\bf Bottom:}
			Noise on the 21-cm GS as a function of $z$.
			The black and red-dashed lines represent the cosmic-variance (CV) noise that we calculate for our fiducial model and for the extreme model of Ref.~\cite{Kaurov:2018kez}, respectively.
			For comparison, we also show the instrumental noise for a GS experiment observing for $t_{\rm obs}=1$ yr as blue lines.
			The upper (dash-dotted) line assumes a bandwidth $B=0.4$ MHz, whereas the lower (dotted) line has a variable $B=2-4$ MHz chosen to produce a comoving width $\Delta \chi=60\,\rm Mpc$, where the cosmic covariance between bins is halved.
		}	
		\label{fig:sigma21}
	\end{figure}

	We show the resulting cosmic variance in Fig.~\ref{fig:sigma21}, along with the global signal for our fiducial parameters.
	The size of the cosmic variance tracks the amplitude of 21-cm fluctuations, which grows at the beginning of cosmic dawn ($z\approx 25$), due to the sourcing of Lyman-$\alpha$ photons, and nearly vanishes during the transition from the Lyman-$\alpha$ coupling to X-ray heating ($z\approx 20$), where the 21-cm global signal reaches a minimum.
	Likewise, the fluctuations grow during the epoch of X-ray heating, and turn around as the entire cosmos is heated (by $z\approx 12$).
	Finally, the EoR sees another growth of fluctuations, as the hydrogen becomes inhomogeneously ionized, and eventually both the 21-cm GS and the fluctuations disappear by $z\approx 6$.

	The cosmic variance for our fiducial model reaches values of $\sigma_{21}\approx 0.05$ mK.
	This is to be compared with the instrumental noise for an experiment targeting the 21-cm GS.
	We find this noise with the standard radiometer equation~\cite{Pritchard:2010pa},
	\be
	\sigma_{\rm inst} (z) = \dfrac{T_{\rm sky}(z)}{\sqrt{B t_{\rm obs}}},
	\label{eq:sigmainst}
	\ee
	where $B$ is the experimental bandwidth, $t_{\rm obs}$ the total observation time, and $T_{\rm sky}$ the sky temperature, dominated by foregrounds.
	We take this last quantity to be $T_{\rm sky} = a_0(\nu/\nu_0)^{-2.5}$, with $a_0=1570$ K at $\nu_0=72$ MHz, in order to match the EDGES data~\cite{Bowman:2018yin}.
	We show, in Fig.~\ref{fig:sigma21}, the instrumental noise for a standard GS experiment with $t_{\rm obs}=1$ year, and $B=0.4$ MHz, as that is the resolution of the public EDGES data.
	Additionally, we show the noise for broader bins, designed to span a comoving distance of 60 Mpc, as we will show later that is the typical correlation length of the GS cosmic variance.
	Those bins have variable widths as a function of redshift, ranging from $B=4$ MHz at $z=6$ to 2 MHz at $z=27$, and produce a noise comparable in size to the cosmic variance for our fiducial case (and we note that the cosmic-variance noise is roughly independent of the bandwidth as long as $\Delta \chi \ll \chi$).

	As clear from Eq.~\eqref{eq:sigma21general}, the size of the CV noise grows with the amplitude of the 21-cm power spectrum,  which is as of yet unmeasured, so models with more marked fluctuations will exhibit larger cosmic variance.
	As an example, we calculate the cosmic variance that would arise in the model of Ref.~\cite{Kaurov:2018kez}, where the parameters of the first galaxies are modified to match the timing of the claimed EDGES detection (albeit not its depth). 
	We show their global signal in Fig.~\ref{fig:sigma21} along ours, which evolves very rapidly during cosmic dawn, as reported by EDGES~\cite{Bowman:2018yin}.
	This produces dramatic 21-cm fluctuations, two orders of magnitude larger than our fiducial model~\cite{Kaurov:2018kez}.
	As a consequence, the expected cosmic-variance noise is much larger, which we compute and show in Fig.~\ref{fig:sigma21}, and can become comparable to the instrumental noise,
	showcasing the importance of including cosmic variance in the analysis of the 21-cm GS.
	We note that, as before, we have fitted the low-$k$ part of the power spectrum to follow matter fluctuations, although for this model we do not have all the low-$k$ data to establish if this was a good fit.
	In addition, this model only fits the timing of the EDGES signal, and the power spectrum would be a factor of 6 larger if the EDGES anomalous depth was confirmed~\cite{Kaurov:2018kez}.

	So far we have focused on the cosmic variance of the 21-cm GS at individual redshifts.
	Nevertheless, cosmic variance will also induce correlations between measurements of the 21-cm GS at nearby redshifts, as those are coherently affected by the same long-wavelength fluctuations.
	As opposed to instrumental noise, this will give rise to a nondiagonal covariance matrix (see Refs.~\cite{Liu:2012xy,Tauscher:2020wso} for nondiagonal matrices due to foregrounds and beam effects).
	To compute it, we start with Eq.~\eqref{eq:var1}, although evaluated at two different redshifts $z_1$ and $z_2$,
	\begin{align}
		\sigma_{21}^2(z_1,z_2) &= \VEV{ T_{21}^{\rm obs}(z_1)T_{21}^{\rm obs}(z_2) } - \overline{T_{21}}(z_1)\overline{T_{21}}(z_2)
		\label{eq:var2}
	\end{align}
	Now the two $T_{21}^{\rm obs}(z_i)$ signals (and as a consequence the window functions $W_{z_i}$ inside the brackets) can have different supports.
	Again going to Fourier space we obtain a generalization of Eq.~\eqref{eq:sigma21general}, 
	\be
	\sigma_{21}^2(z_1,z_2) = \int \dfrac{d^3k}{(2\pi)^3} P_{21}(k,z_1,z_2) \mathcal W_{z_1}(k) \mathcal W_{z_2}(k),
	\label{eq:sigma21correlated}
	\ee
	where $P_{21}(k,z_1,z_2)$ is the power spectrum of 21-cm fluctuations at $z_1$ and $z_2$, which we describe in Appendix~\ref{app:covariance}.
	In order to build some intuition let us show what this integral looks like for two adjacent slices of our cosmos, centered at $\chi$ and $\chi+\delta \chi$ (or $z$ and $z+\delta z$ in redshift).
	There we can approximate
	\be
	\mathcal W_{z+\delta z}(k) = j_0\left [k \chi+\delta \chi\right] \approx  \mathcal W_z(k) \cos(k \delta \chi),
	\ee
	for $\delta \chi \ll \chi$.
	Under that approximation it is clear that the cosmic covariance between redshifts will be suppressed for large separations $\delta z$, although, as expected, 21-cm fluctuations with small $k$ will correlate slices that are roughly as far as $\delta \chi\sim k^{-1}$.

\begin{figure}[btp!]
		\centering
		\includegraphics[width=0.48\textwidth]{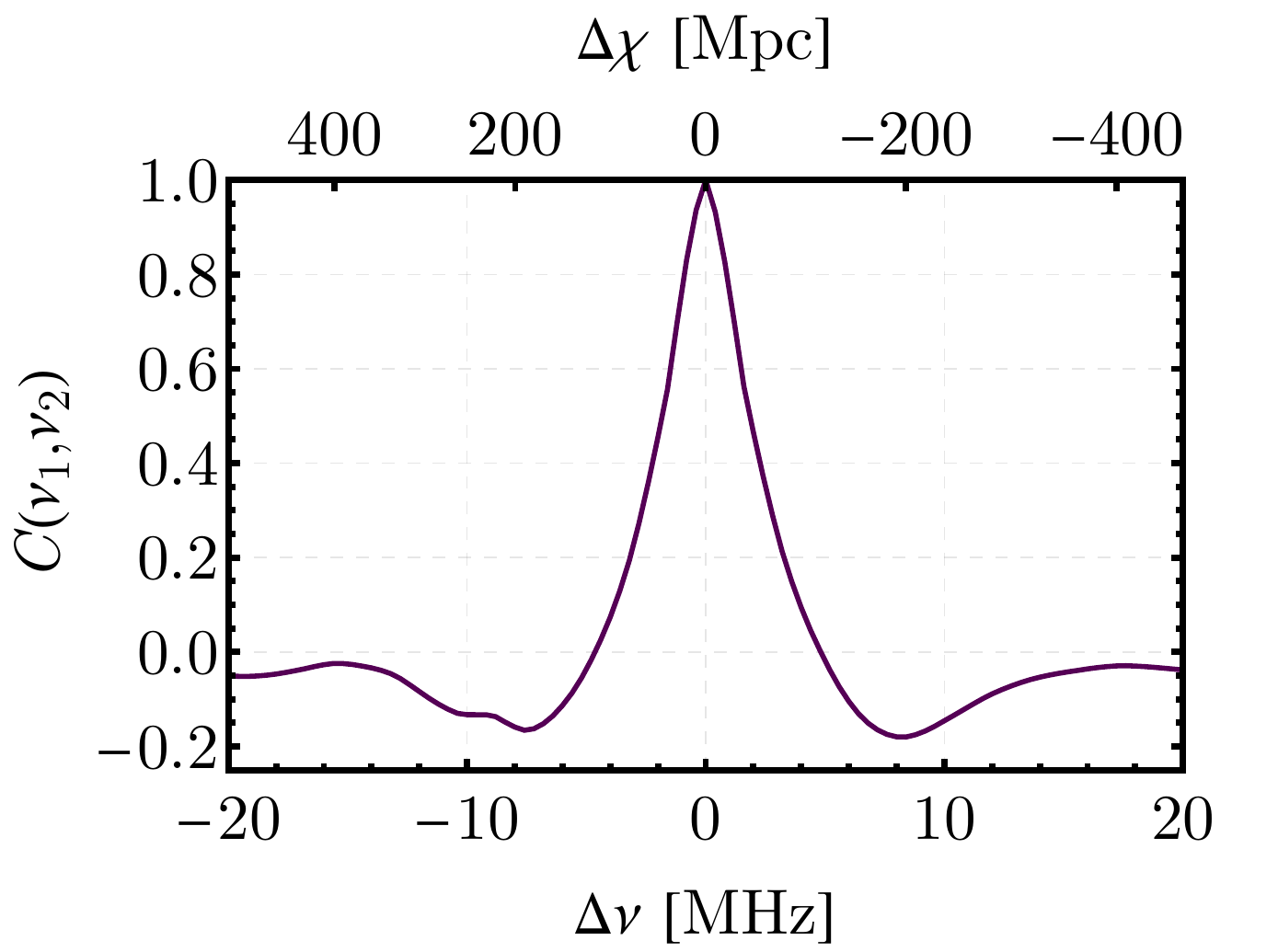}
		\caption{Normalized correlation, $C(\nu_1,\nu_2) = \sigma_{21}^2(\nu_1,\nu_2)/[\sigma_{21}(\nu_1)\sigma_{21}(\nu_2)]$, between the 21-cm GS measured at $\nu_1=80$ MHz ($z_1=16.8$) and other frequencies $\nu_2=\nu_1+\Delta \nu$, separated by multiples of $0.4$ MHz.
			In the top $x$ axis we mark the comoving distance between frequencies, where positive numbers move upwards in redshift.
			Slices up to $\Delta \nu\sim 10$ MHz (or $\Delta\chi\sim 200$ Mpc) are correlated with each other, although the correlation drops by half by $\chi_{\rm corr}=60$ Mpc.
		}	
		\label{fig:Corr21fixz}
	\end{figure}

	We show, in Fig.~\ref{fig:Corr21fixz}, the (normalized) correlation between measurements of the 21-cm GS at $80$ MHz ($z=16.8$), within the band of most GS experiments, and other frequencies.
	Nearby measurements are positively correlated, whereas for displacements $\Delta \nu \approx 10$ MHz (or $\Delta \chi\approx 200$ Mpc) the correlation becomes slightly negative, and vanishes at infinity.
	Displacements of $\chi_{\rm corr}\approx 60$ Mpc are sufficient to halve the correlation, roughly independently of the central redshift.
	We show the covariance between all redshifts in Appendix~\ref{app:covariance}.

	The cosmic variance that we have calculated acts as an additional noise term in the GS covariance matrix, which has several effects.
	First, cosmic-variance noise will decrease the significance of any detection, by increasing the error budget.
	We find that for our fiducial 21-cm model, and a year-long campaign to detect the GS this is only a percent-level effect, although for an extreme model (as the one from Ref.~\cite{Kaurov:2018kez} presented above) it produces a degradation of 70\%.
	Further, we find that for our model the cosmic-variance limit, with no instrumental noise, boasts a signal-to-noise ratio $\approx 10^4$, which albeit very large is finite.
	Second, the inferred parameters of the 21-cm model will have underestimated errors. This underestimation again ranges from 1\% for our case with mild fluctuations to nearly 100\% for the extreme case.
	Furthermore, a detection of cosmic variance would open the door to an indirect measurement of the 21-cm fluctuations integrated over low-$k$.
	We expand on these results in Appendix~\ref{app:chisquared}.

	As hinted above, a determinant factor for the size of the large-scale 21-cm fluctuations---and thus the cosmic-variance noise---is how quickly the global signal evolves.
	This allows for a heuristic calculation of the cosmic variance with the simple shape $\sigma_{21,\rm approx} (z) = a \times d\overline{T_{21}}/dz$, with an amplitude $a\simeq10^{-3}$ which we calibrate in Appendix~\ref{app:approx}, where we also show how well this approximation compares with the full integrals computed above.

	While we have focused on the cosmic variance of the 21-cm GS in isolation, the same effect will create a cross-correlation between the 21-cm GS and the power spectrum (see, e.g.~\cite{Krause:2016jvl} for an application to the large-scale structure). 
	Thus, joint analyses of the 21-cm GS and power spectrum, as proposed in, e.g., Ref.~\cite{Liu:2015gaa}, ought to include cosmic covariance.

	As a byproduct of this work we have performed the largest cosmic-dawn and EoR simulations to date (although not the highest-resolution ones, e.g.~\cite{Gnedin:2014uta,Ocvirk:2015xzu,Poole:2015mhx,Mesinger:2016ddl}), with a box size of $L=1.8$ Gpc comoving in {\tt 21cmvFAST}.
	Such large box sizes were required to find the long-wavelength behavior of the 21-cm fluctuations, which determines the size of the cosmic variance, as well as the correlation between bins.
	This has provided clarity about the small-$k$ behavior of the 21-cm power spectrum.
	We emphasize, nonetheless, that the cosmic-variance effect presented here does not rely on the details of the algorithm in {\tt 21cmvFAST}/{\tt 21cmFAST}, and could be computed with any other simulated or analytic power spectrum.
	Moreover, while we have computed the cosmic variance using analytic methods, we show in Appendix~\ref{app:simulations} that our formalism agrees with the direct variance observed in simulations.

	In summary, the 21-cm GS suffers from cosmic variance, similar to other cosmological observables.
	In this {\it Letter} we have presented this effect in detail for the first time, and computed its size.
	While it is unlikely to hamper a first detection of the 21-cm GS, cosmic variance provides an irreducible source of noise that has to be taken into account.
	Doing so brings us one step closer to understanding cosmic dawn and the epoch of reionization at the percent level.
	\\
	
	\acknowledgements    
	We are thankful to Alexander Kaurov and Andrei Mesinger for discussions.
	JBM is funded by NSF grant AST-1813694.

	\bibliography{cv21cm}

	\newpage
	
	\appendix

	\section{Simulation Specifics}
	\label{app:simsspecs}

	In this work we model the 21-cm signal using {\tt 21cmvFAST}/{\tt 21cmFAST}\footnote{\url{https://github.com/JulianBMunoz/21cmvFAST.} \\ 
		\url{https://github.com/andreimesinger/21cmFAST.}}, a semi-numerical simulation package that accounts for the formation of the first stars (including the effect of densities and streaming velocities), as well as the effects of Lyman-$\alpha$, X-ray, and ultraviolet (UV) ionizing photons.
	We perform simulations with a box size of $L=1.8$ Gpc (all cosmological distances are comoving unless otherwise noted), which are large enough to resolve all the relevant scales for our application.
	We used the best-fit cosmological parameters from Planck~\cite{Aghanim:2018eyx}.
	For the astrophysical parameters of {\tt 21cmvFAST} we have set a stellar fraction of $f_*=0.1$, and $T_{\rm vir}=10^{4.3}$ K for UV sources, and a time-dependent Lyman-Werner feedback for the X-ray and Lyman-$\alpha$ sources, implemented as in Refs.~\cite{Munoz:2019fkt,Munoz:2019rhi}.
	We take an X-ray luminosity of $\log_{10}(L_X/{\rm SFR})=40$, where SFR is the star-formation rate, with a log-flat spectrum over the $0.2-2$ keV energy range.
	For the UV part we take an ionizing efficiency $\zeta=20$ and a mean-free path of $R_{\rm mfp}=15$ Mpc~\cite{Mesinger:2010ne}.

	\section{Selection Functions}
	\label{app:windows}
	
	In this section we lay the technical formalism for our selection functions, as well as some alternatives to it.

	Throughout the text we assume that a GS experiment observes the entire sky, over a narrow (but non-zero) width $\Delta \chi$, at each redshift $z$, corresponding to a comoving distance $\chi$ from us.
	In that case we can simply model the window function as
	\be
	W_z(\Omega,\chi) = \dfrac 1 { {\chi}^{2} \Delta \chi} \Theta(\chi_{\rm max}-\chi) \Theta(\chi-\chi_{\rm min}),
	\label{eq:tophat}
	\ee
	where $\Omega$ is the solid angle, and $\chi_{\rm max/min}={\chi} \pm \Delta \chi/2$.
	where $\chi(z)$ is the comoving distance to redshift $z$.
	The Fourier transform of this function is found to be
	\be
	\mathcal W_z(k)= \int_{\chi_{\rm min}}^{\chi_{\rm max}}  \dfrac{d \chi' \chi'^2}{\chi^2\Delta \chi} j_0(k \chi') \approx j_0(k \chi)
	\ee
	where $\chi_{\rm max/min}=\chi \pm \Delta \chi/2$, whereas the last approximation---which we have used throughout the main text---is valid for $\Delta \chi \ll \chi$ and $k<1/\Delta \chi$.
	The former will always be true, and the latter will hold for the relevant $k$ range where the integral has weight.

	\subsection*{Spherical Harmonics}
	
	We have chosen to work in $k$-space as that is the most common language for the 21-cm fluctuations.
	Alternatively, we could have phrased our formalism in terms of spherical harmonics (with multipoles $\ell$ and $m$), as usually done for instance in  CMB analyses.
	In that case we would say that over a single (not necessarily narrow) slice at redshift $z$
	\be
	T_{21}(\Omega,z) = \overline{T_{21}}(z) + \sum_{\ell,m} a_{\ell,m}(z) Y_{\ell,m}(\Omega).
	\ee
	Then, what in the main text we called the observed global signal, which is a single realization of this field, is
	\be
	T_{21}^{\rm obs}(z) = \overline{T_{21}}(z) + a_{00}(z).
	\ee
	While this last monopole term has a zero expected value, it will fluctuate. 
	The variance of $a_{00}$ is given by
	\be
	C_0 = (4\pi)^2 \int dk k^2 P_{21}(k) |w_0(k \chi)|^2,
	\ee
	where $w_\ell(x)=j_\ell(x)$ is the usual geometric factor of each $\ell$.
	Then, the variance of the 21-cm global signal is just
	\be
	\sigma_{21}^2 = \dfrac{C_0}{4\pi} = \int \dfrac{d^3k}{(2\pi)^3} P_{21}(k) \left |j_0( k \chi) \right|^2,
	\ee
	as we found in Eq.~\eqref{eq:sigma21general} of the main text.
	
	We note in passing that a similar variance affects the CMB monopole temperature, $T_0\simeq2.725$K. There, the leading contribution to this variance is from the Sachs-Wolfe effect \cite{1967ApJ...147...73S}
	\be
	\sigma_{\rm CMB}^2\approx \int dk\, k^2 P_\Phi(k)\left[\frac{\Psi(k,z_*)}{3\Phi(k,0)}\right]^2 \left |j_0( k \chi(z_*)) \right|^2,
	\ee
	where $P_\Phi(k)$ is the primordial spectrum of fluctuations, $\Phi$ and $\Psi$ are gravitational potential in conformal Newtonian gauge, and $z_*$ is the redshift of recombination. Integrating over all perturbations within our Hubble patch today leads to 
	$\sigma_{\rm CMB}\approx 50\,\rm \mu K$. This is to be compared with the current instrumental error in $T_0$ of $570 \,\mu$K~\cite{Fixsen:2009ug}.

	\subsection*{Flat-sky}
	
	We now explain how the calculation would be performed in the flat-sky approximation, commonly assumed for large-scale structure surveys, and compare the results.
	In the flat-sky limit we assume that we observe a small patch of the sky over an angle $\theta_S$.
	In that case the window function is anisotropic, and can be written as
	\be
	W_z(\mathbf x) = \dfrac 1 {\pi {\chi}^{2} \Delta \chi} \Theta(\chi_{\rm max}-\chi_{||}) \Theta(\chi_{||}-\chi_{\rm min}) \Theta(\theta_S{\chi}-\chi_\perp ) ,
	\ee
	where the subindices $||$ and $\perp$ mean line-of-sight and perpendicular. 
	Then, its Fourier transform is
	\be
	\mathcal W_z(\mathbf k) = \int dx_\perp x_\perp \int dx_{||}  W(\mathbf x) = \mathcal W_z^\perp(k_\perp) \mathcal W_z^{||}(k_{||}),
	\ee
	where the two window functions are given by~\cite{Krause:2016jvl}
	\be
	\mathcal W_z^{||}(k_{||}) = j_0[(k_{||} {\chi})/2],
	\ee
	and
	\be
	\mathcal W_z^{\perp}(k_{\perp}) = \dfrac{2 J_1(k_{\perp} {\chi} \theta_S)}{k_{\perp} {\chi} \theta_S}.
	\ee
	
	We find that using the flat sky appriximation results in a 21-cm cosmic variance that is a factor of 3 larger than the full-sky case.
	This is perhaps to be expected, as for the full sky (or even half of it) $\theta_S\sim 1$.
	Therefore, there are no modes parallel to the line of sight, as it varies by $\sim \pi$ across the observation region.

	\section{Modeling the large-scale 21-cm fluctuations}
	\label{app:largescale}

	Here we explain how we model the large-scale 21-cm fluctuations using our simulations.
	
	For notational convenience let us define the amplitude of fluctuations of quantity $i$ as
	\be
	\Delta^2_{i}(k)  = \dfrac{k^3}{2\pi^2} P_{i}(k),
	\ee
	where $P_i$ is its power spectrum, although in this Appendix we will often refer to $\Delta^2_i$ as the power spectrum unless confusion can arise.

	\begin{figure}[btp!]
		\includegraphics[width=0.52\textwidth]{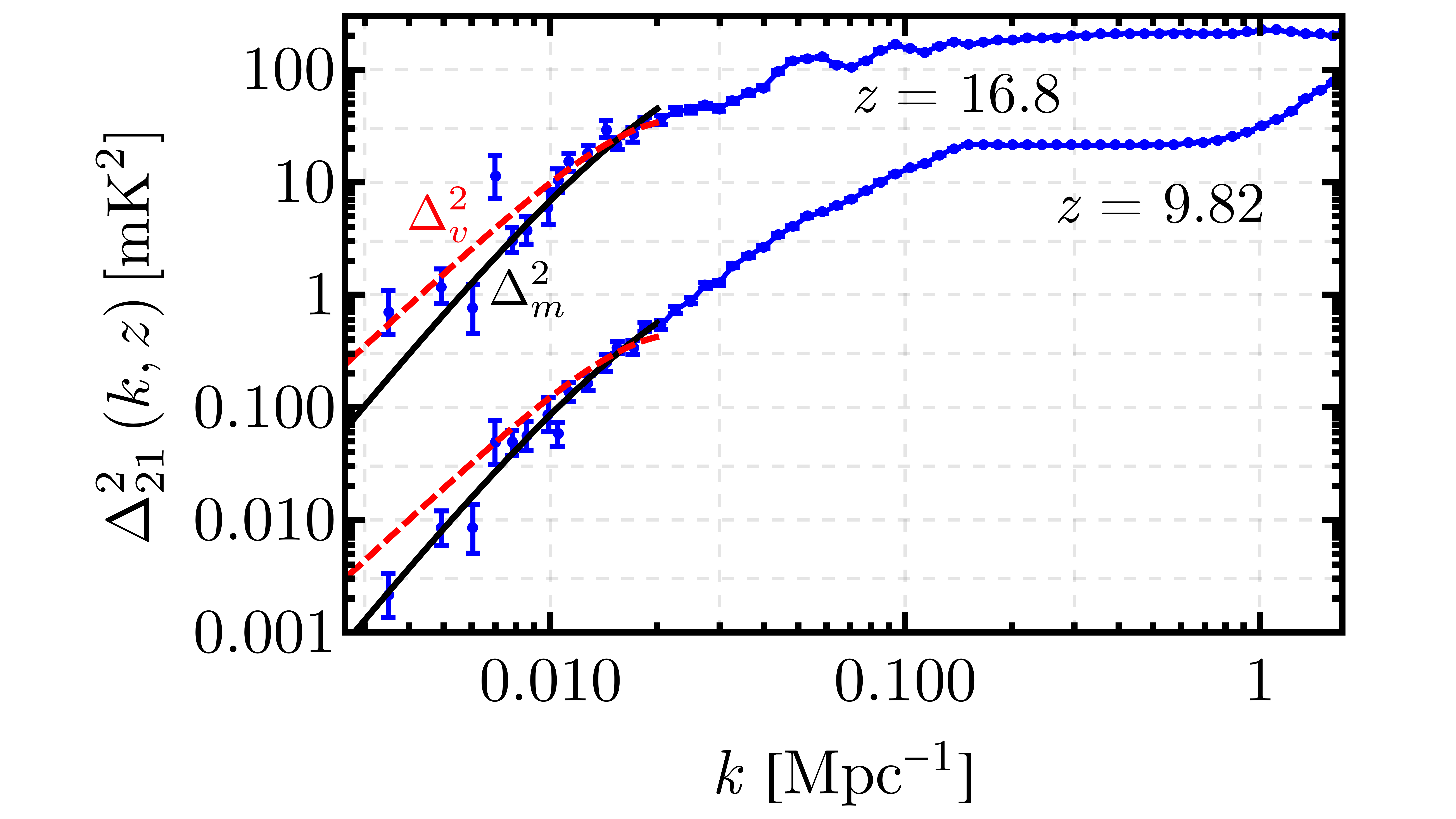}
		\caption{Simulated amplitude of 21-cm fluctuations as a function of wavenumber $k$ at $z=9.82$ and $z=16.8$, obtained with {\tt 21cmvFAST}.
			The black and red-dashed lines show the two assumptions for the low-$k$ behavior of the 21-cm power spectrum, either following the matter fluctuations ($\Delta^2_m$) or the relative-velocity ones ($\Delta^2_v$).
			These assumptions provide comparable fits to the simulation output during cosmic dawn $(z>12)$, although matter fits better during the EoR ($z<12$). Nevertheless, both yield very similar overall results.
		}	
		\label{fig:P21}
	\end{figure}

	We show in Fig.~\ref{fig:P21} the 21-cm  power spectrum at $z=9.82$ (during the EoR) and 16.8 (during cosmic dawn), for one of our large-box ($L_{\rm box}=1.8$ Gpc) {\tt 21cmvFAST} simulations.
	As it is clear from this figure, the low-$k$ wavenumbers have larger error bars when measured in simulations, as there are fewer modes per $k$.
	Nevertheless,  theoretically we expect that at large scales (low $k$) the 21-cm fluctuations trace the matter and relative-velocity fluctuations with some overall bias coefficients, since at sufficiently large scales the fluctuations ought to be linear~\cite{McQuinn:2018zwa} (as is the case for the large-scale structure~\cite{Bernardeau:2001qr}).
	Then, we can write
	\be
	\Delta^2_{21}(k,z) = b_m^2(z) \Delta^2_{m}(k,z) + b_v^2(z) \Delta^2_{v}(k,z),
	\ee
	where $\Delta^2_{m}$ is the matter power spectrum,
	and $\Delta^2_{v}$ is the power spectrum of the DM-baryon relative velocities.
	Fig.~\ref{fig:P21} shows that this expression only holds for very large scales ($k \lesssim 0.02 \,\Mpcinv$), where we can ignore the non-linearity of the first stellar formation.
	For such low $k$ we will assume that $\delta T_{21}$ traces only matter or velocity fluctuations, for simplicity (and because the final results are similar). Equivalently, we will assume that either $b_m$ or $b_v$ are zero.

	\subsection*{Matter bias}
	
	Let us begin by assuming that $b_v=0$, and thus $\delta T_{21} = b_m \delta_m$.
	We use the modes with $k<0.02$ Mpc$^{-1}$ to find $b_m(z)$ simply by fitting our simulation results at each redshift.
	For reference, $b_m$ grows during cosmic dawn, reaching a peak during the Lyman-coupling era, another during the epoch of heating, and finally a smaller one during the EoR,
	becoming lower both at higher and lower $z$, as well as in the transition between the two eras, as expected of the overall large-scale fluctuations~\cite{Munoz:2019rhi}.
	
	This is the assumption we take in the main text.
	Our approach is similar to that proposed in Ref.~\cite{Somerville:2003bq} for finding the cosmic-variance error in galaxy counts, where the variance of matter fluctuations was first calculated, and then multiplied by a bias coefficient.
	Here, however, we cannot calculate the bias from first principles, so we use simulation results.

	\subsection*{Velocity bias}

\begin{figure}[hbtp!]
		\includegraphics[width=0.52\textwidth]{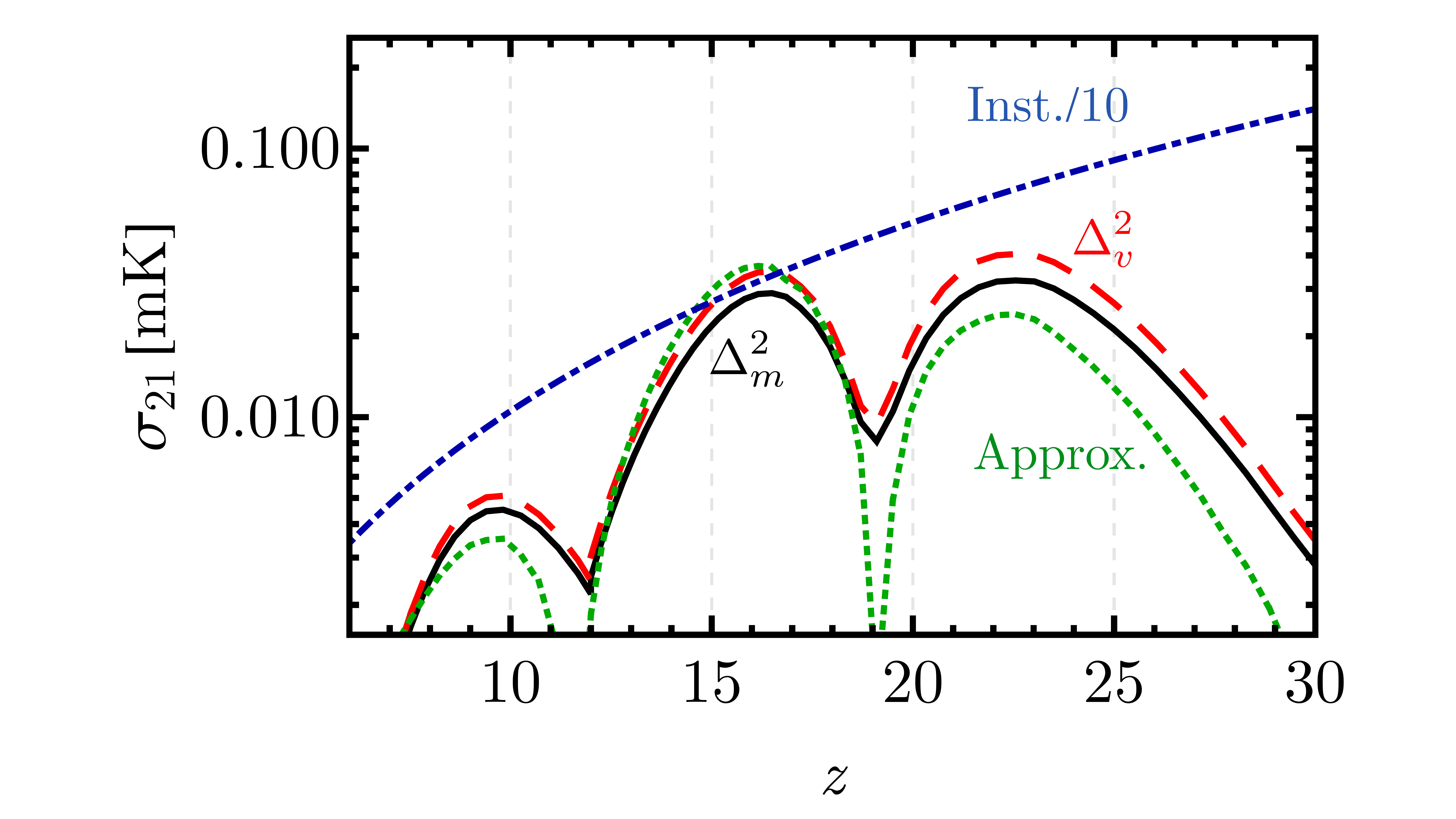}
		\caption{Different noises on the 21-cm GS as a function of redshift.
			The blue dash-dotted line shows the instrumental error, divided by ten, and the black line the cosmic variance in our fiducial model, as in Fig.~\ref{fig:sigma21}.
			The green-dotted line shows our approximate calculation of the cosmic-variance noise, explained in Appendix~\ref{app:approx}, which closely follows the exact result in black, and the red-dashed line shows the result we would have obtained if we assumed the 21-cm power spectrum follow the relative-velocity power spectrum ($\Delta^2_v$) instead of the matter one ($\Delta^2_m$).
		}	
		\label{fig:sigma21approx}
	\end{figure}

	We now study the alternative case, where $b_m=0$ and the 21-cm fluctuations trace the DM-baryon relative velocities.
	We get similar results as with matter, as shown in Fig.~\ref{fig:sigma21approx}.
	We note that while $\Delta^2_v$ and $\Delta^2_m$ provide comparably good fits during cosmic dawn (as both are tracers of the 21-cm signal), the same is not true for the EoR, where the 21-cm power spectrum tracks $\Delta^2_m$ much closer, as is clear for the $z=9.82$ power spectrum in Fig.~\ref{fig:P21}.
	This is to be expected, as in our implementation of {\tt 21cmvFAST} the relative velocities modulate the amount of X-ray and Lyman-$\alpha$ photons emitted, but not the UV photons responsible for reionization.
	
	We note that cosmic variance was previously neglected in Ref.~\cite{Liu:2012xy} as they assumed that there were no 21-cm fluctuations on scales larger than $\sim 1$ deg. (or $k\lesssim 10^{-2} \,\Mpcinv$).
	Our large-box {\tt 21cmvFAST} simulations show that there are 21-cm fluctuations at very large scales, down to $k\sim 10^{-3}\,\Mpcinv$,
	which has allowed us to compute the cosmic variance of the 21-cm GS for the first time.

	Assuming that $\Delta_{21}^2$ follows either the matter or velocity power spectrum allows us to integrate down to arbitrary wavenumbers.
	We have tested that setting the lower limit of the integrals below the simulation cutoff $k_{\rm min}=2\pi/L_{\rm box}=3\times 10^{-3}\,\Mpcinv$ only increases the cosmic variance by 3\%, showing that our boxes are large enough to capture the effect.
	To further illustrate this point,
	we compute the ``theoretical error" that we would incur when finding the 21-cm GS in a box of size $L_{\rm box}$.
	We do so by performing the integral
	\be
	\sigma_{21,\rm th}^2 = \int_{k_{\rm min}}^{k_{\rm box}} \!\! \dfrac{dk}{k}\, \Delta^2_{21}(k)
	\ee
	for $k_{\rm min}=3\times 10^{-3}\,\Mpcinv$, as before, and $k_{\rm box}=2\pi/L_{\rm box}$ depends on the box size.
	With this formula we find that the theoretical noise is approximately $\sigma_{21,\rm th}\approx 1\,{\rm mK} \times (L_{\rm box}/10^3\,\rm Mpc)^{-1}$, at $z=16.8$.
	In particular, we find that the finite box produces noises of $\sigma_{21,\rm th}=\{1,2,3,10\}$ mK for $L_{\rm box}=\{10^3,600,400,100\}$ Mpc, which are commonly assumed. 
	Consequently, we strongly recommend the use of boxes with sizes larger than 600 Mpc whenever possible, not only to properly account for photon propagation and model the power spectrum~\cite{Kaur:2020qsa}, but also the global signal.
	We note that this accounts for statistical uncertainty, and would be added to any systematic theoretical errors from mismodeling the signal.

	\section{Cosmic Covariance Matrix}
	\label{app:covariance}

	Here we define how we calculate the cosmic covariance matrix.
	We begin by defining the 21-cm power spectrum at two redshift bins, $P_{21}(z_1,z_2)$.
	This quantity can, in principle be computed from simulations, by taking the two point function in Fourier space of {\tt 21cmvFAST} boxes at $z_1$ and $z_2$.
	This is, nevertheless, computationally costly, so we follow a simpler approach. Under the assumption that the large-scale 21-cm fluctuations follow the matter overdensities with a bias $b_m$, we can write
	\be
	P_{21}(k,z_1,z_2) = b(z_1)b(z_2)\Delta^2_m(k,z_1)D_+(z_2)/D_+(z_1),
	\ee
	for $k< 0.02\,\Mpcinv$, where $D_+(z)$ is the linear growth factor. 
	As before, we will interpolate from our simulation output for $k\geq 0.02\,\Mpcinv$, for which this formula will not hold.
	There we simply take $P_{21}(k,z_1,z_2)=\sqrt{P_{21}(k,z_1)P_{21}(k,z_2)}$, which does not affect our results significantly, as the majority of the integral weight is at lower $k$.

\begin{figure}[hbtp!]
		\centering
		\includegraphics[width=0.48\textwidth]{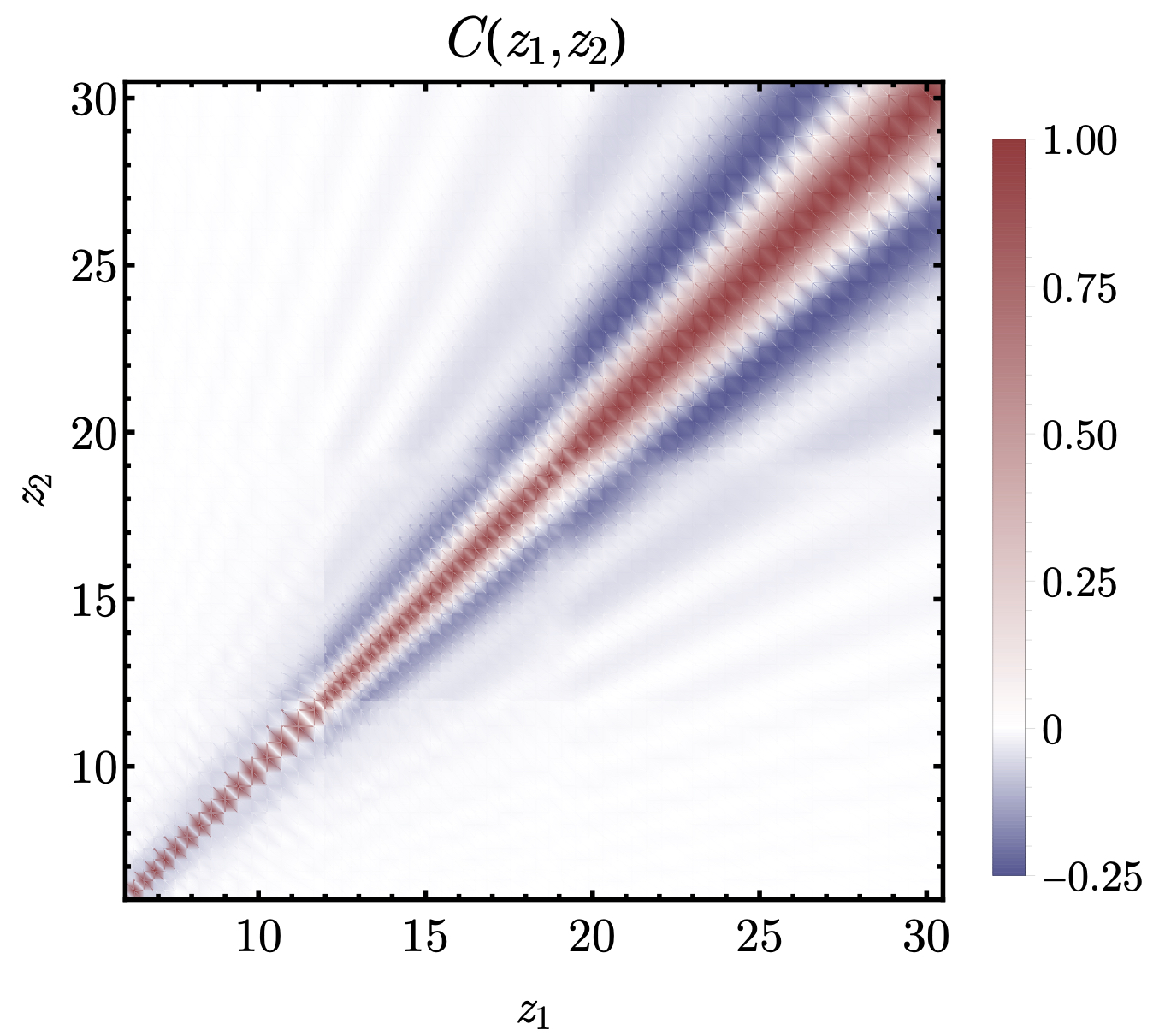}
		\caption{Normalized correlation $C(z_1,z_2)=\sigma_{21}^2(z_1,z_2)/[\sigma_{21}(z_1)\sigma_{21}(z_2)]$ between 21-cm slices at different redshifts $z_1$, $z_2$, due to cosmic variance.
			Large-scale fluctuations induce important ($\mathcal O(1)$) correlations even outside the diagonal.
		}	
		\label{fig:Corr21}
	\end{figure}

	We show the normalized correlation between redshifts, defined as
	\be
	C(z_1,z_2)=\dfrac{\sigma_{21}^2(z_1,z_2)}{\sigma_{21}(z_1)\sigma_{21}(z_2)},
	\ee
	in Fig.~\ref{fig:Corr21}, where the variances have been computed with the $P_{21}(k,z_1,z_2)$ delineated above.
	This figure illustrates how cosmic variance induces correlations between different redshifts that are far from each other.
	We warn the reader that the finite redshift resolution of {\tt 21cmvFAST} slices is $\Delta z \approx 0.3$, which corresponds to a frequency difference of $\approx 1$ MHz. Thus, correlations between bins with $\Delta \nu<1$ MHz are interpolated from our results, and should be interpreted with caution.
	
	\section{Impact of cosmic variance on sensitivity}
	\label{app:chisquared}
	
	In this Appendix we explore how much the cosmic-variance noise that we have calculated in this work affects the sensitivity of different tests done with the 21-cm GS.

	\subsection*{Signal to Noise}

	First, we estimate the signal-to-noise ratio (SNR) with and without cosmic variance.
	This is calculated simply through
	\be
	{\rm SNR}^2 = \sum_{i,j} d_i \mathcal C_{ij}^{-1} d_j,
	\label{eq:SNR}
	\ee
	where the indices $i$ and $j$ run over each frequency channel, $d_i$ is the vector that contains the 21-cm global signal (not including foregrounds), and $\mathcal C^{-1}$ is the inverse of the covariance matrix.
	The full covariance matrix is defined to be
	\be
	\mathcal C_{ij,\rm full} = \delta_{ij} \sigma_{\rm inst}^2(\nu_i) + \sigma_{21}^2(\nu_i,\nu_j),
	\label{eq:Covtot}
	\ee
	although we will also compute quantities assuming no cosmic variance, in which case only instrumental noise contributes to the covariance matrix, and $\mathcal C_{ij,\rm inst} = \delta_{ij} \sigma_{\rm inst}^2(\nu_i)$, where $\sigma_{\rm inst}$ is defined in Eq.~\eqref{eq:sigmainst}.
	Note that the SNR does not take into account any marginalization over foregrounds, although it will suffice to indicate the effect of mismodeling the covariance matrix.
	We will assume an experiment observing for a year with 0.4 MHz bandwidth, covering the range  50-200 MHz (corresponding to $z=6-27$, which for instance can represent EDGES low and high bands~\cite{Monsalve:2018fno}).
	In that case, the SNR of our fiducial 21-cm signal is SNR $=2320$ when assuming only instrumental noise, versus $2284$ when including the full covariance matrix (a 2\% reduction).

	As mentioned in the main text, the cosmic-variance noise grows with the amplitude of the 21-cm power spectrum, which is as of yet unmeasured.
	To account for variations around our fiducial, we will rescale the cosmic-variance part of Eq.~\eqref{eq:Covtot} by different factors to showcase how large this effect could be.
	In our fiducial model the typical 21-cm fluctuations have a size of $\Delta_{21}\sim10\,\rm mK$, whereas the GS peaks at $\overline{T_{21}}\approx -100$ mK, so $\mathcal O(1)$ fluctuations would produce a power spectrum $\Delta^2_{21}\sim10^4\,\rm mK^2$, two orders of magnitude larger than we consider (and comparable to those in Ref.~\cite{Kaurov:2018kez}).
	To account for this possibility, we calculate results rescaling our 21-cm power spectrum by a factor of 10 and 100, where we find that  the SNR is reduced to 2128 with a 10-fold increase, and further down to 1594 with a a 100-fold increase.
	These are, respectively, 10\% and 70\% lower than in the case without cosmic variance.
	These results show that cosmic variance is expected to reduce the overall SNR, by a few percent for a standard 21-cm model, and up to $\mathcal O(1)$ for more extreme cases.

	As a theoretical exercise, we now consider a cosmic-variance limited experiment.
	In that case there would be no instrumental noise, and the cosmic variance we calculated is the only component of $\mathcal C_{\rm full}$ in Eq.~\eqref{eq:SNR}.
	We find that the maximum SNR achievable for our fiducial model is SNR $=1.6\times 10^4$,  roughly an order of magnitude larger than the case with instrumental noise presented above.
	We have only computed this quantity for large bandwidths $B\gtrsim$ MHz, as for smaller $B$ the correlation between nearby bands has to be interpolated between our simulation snapshots, which are separated by roughly 1 MHz, as explained in Appendix~\ref{app:covariance}.
	While this cosmic-variance-limited SNR is very large, it shows that there is a fundamental limit to how well the 21-cm GS can be measured, even with an arbitrarily precise instrument.
	We note, however, that if the power spectrum was a factor of 100 larger, as discussed above, the maximum achievable SNR would be 160 when integrated across all frequency bins, setting a fairly low ceiling for the cosmic-variance limit.

	Here we have only assumed instrumental noise to compute the SNR.
	Nevertheless, independent analyses have found that the noise in EDGES data can be as large as 30 mK~\cite{Hills:2018vyr,Bradley:2018eev}, so $\sigma_{\rm inst}$ may be influenced by systematic effects in addition to thermal noise, which we do not consider in this work.
	In addition, both foregrounds and beam effects can also produce off-diagonal elements in the covariance matrix~\cite{Liu:2012xy,Tauscher:2020wso}.

	\subsection*{Parameter Errors}

	Now we estimate whether a prospective cosmological detection of 21-cm can be extracted from foregrounds including cosmic variance, and how well we can know the timing and depth of that signal.
	In order to make progress, we will generate mock data by taking our model for the 21-cm GS and adding foregrounds and random errors drawn from the full covariance matrix, which includes both instrumental and cosmic-variance errors.

	Instead of following an effective model, such as the flattened Gaussian of Ref.~\cite{Bowman:2018yin}, we will simply take our model to follow our GS ($\overline{T_{21}}(z)$), with an arbitrary amplitude and shift in redshift.
	We then write our model for the GS as
	\be
	\overline{T_{21}^{\rm mod}}(z) = A_{21} \overline{T_{21}}(\alpha_{21} z),
	\ee
	and attempt to measure the two parameters $A_{21}$ and $\alpha_{21}$, with fiducial values of unity. 
	In the spirit of simplicity, we will only include a single foreground component following $T_{\rm fore} = a_0 (\nu/\nu_0)^{-2.5}$, with an amplitude $a_0$ that we simultaneously fit for.
	
	Given our three model parameters ($A_{21}$, $\alpha_{21}$, and $a_0$), we calculate the $\chi^2$ of our model against the mock data through
	\be
	\chi^2 = \sum_{i,j} v_i \mathcal C_{ij}^{-1} v_j,
	\label{eq:chisq}
	\ee
	where $v_j$ is the vector carrying the difference between our model and data.
	This expression is related to the log-likelihood under the assumption that errors are Gaussian.

	We start by minimizing the $\chi^2$ both with the full covariance matrix and the instrumental only.
	We find that the minimized $\chi^2$ is larger in the latter case, by $\Delta \chi^2=5$.
	That is because the data is generated from the full covariance matrix, and thus it can fully capture its correlations.
	This difference grows to $\Delta \chi^2=18$ and $\Delta \chi^2=104$ for a 21-cm power spectrum that is a factor of 10 and 100 larger, respectively.

	Ignoring cosmic variance can, additionally, underestimate the error-bars in cosmological parameters.
	We will forecast errors in our two effective parameters, $A_{21}$ and $\alpha_{21}$, in all cases marginalizing over the amplitude $a_0$ of the foregrounds.
	In our fiducial case we find that the forecasted errors (with values of $\sigma(A_{21})=5.0\times 10^{-4}$ and $\sigma(\alpha_{21})=8.2\times 10^{-5}$) are undrestimated by 1\% if cosmic variance is ignored.
	For a 10-fold increase in the 21-cm power spectrum the real error-bars are 10\% larger than those obtained with only instrumental error, whereas for a 100-fold increase they become larger by a factor of 1.3 and 3.7 for $A_{21}$ and $\alpha_{21}$, respectively.
	While the strategy of varying $A_{21}$, $\alpha_{21}$, and $a_0$ is likely not a good approach to analyze real data, it suffices to study how much cosmic variance shifts results.

	An interesting consequence of this simple analysis is that cosmic variance is detectable in data when comparing against instrumental noise only, as we find $\Delta \chi^2\approx 5$ between those two cases using Eq.~\eqref{eq:chisq}.
	If we were able to establish the presence of cosmic-variance noise, for instance with the $\chi^2$ test proposed above, and determine its size, we would learn about the 21-cm fluctuations integrated over low $k$, which cannot be measured directly with interferometers due to foregrounds~\cite{Vedantham:2011mh,Morales:2012kf,Pober:2013ig,Liu:2014bba}.

	\begin{figure}[hbtp!]
		\includegraphics[width=0.52\textwidth]{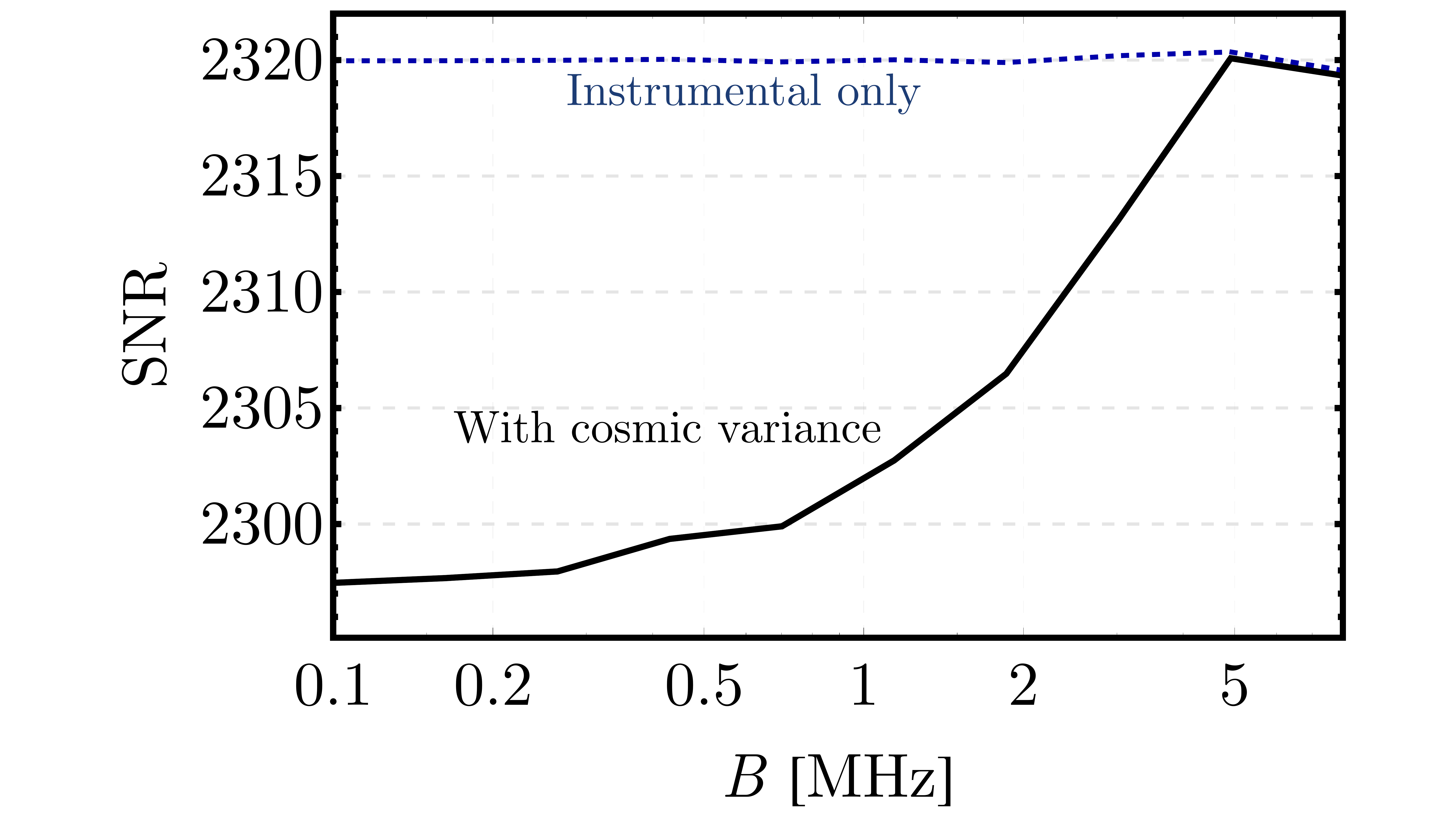}
		\caption{Signal-to-noise ratio (SNR), calculated from Eq.~\eqref{eq:SNR}, as a function of the bandwidth $B$ assumed for the 21-cm GS experiment, in all cases with 1 year of observation.
			The full SNR, including cosmic variance, drops below the instrumental-only result for low $B$, as nearby bins become highly correlated with each other.
			For reference, the typical correlation length of 21-cm fluctuations corresponds to $B\approx 3$ MHz.
		}	
		\label{fig:SNRbins}
	\end{figure}

	\subsection*{As a function of bandwidth}
	
	As a final check in this Appendix, we study how our results would vary when changing the bandwidth (binning) of our data.
	Throughout this work we have mainly assumed a constant bandwidth $B=0.4$ MHz, as that was the value reported by EDGES in their detection.
	Both finer and broader bins are possible, and here we study how the cosmic variance affects the extraction of the 21-cm GS for different bandwidths.
	For that we will use the SNR  as a benchmark of a prospective detection.
	
	Lowering the value of $B$ produces more frequency channels, and thus finer $z$ resolution, albeit at the cost of larger thermal noise.
	If only instrumental noise was present, these two factors would cancel out, yielding SNR $= 2320$, as shown in Fig.~\ref{fig:SNRbins} (where the small wiggles around this value arise from the integer number of bands that can fit over the entire frequency range).
	Including cosmic variance, however, introduces a preferred scale in the problem, as adjacent redshifts are correlated roughly up to a comoving distance $\chi_{\rm corr}=60$ Mpc, corresponding to $B=2-4$ MHz (for $z=6-27$).
	For larger $B$ the cosmic variance has nearly no effect, as even close-by bins are uncorrelated.
	For smaller widths, however, the SNR drops below the instrumental-only curve, as nearby bins contain highly correlated information
	We recover the result from above, where the SNR for $B=0.4$ MHz is approximately 1\% lower when including cosmic variance.
	As before, the size of this decrement in SNR will be larger if the 21-cm power spectrum grows.

	\section{Approximately including cosmic variance}
	\label{app:approx}
	
	In this Appendix we show a simple heuristic way of including the cosmic variance for any 21-cm GS without having to perform any integrals, and in fact without directly using the 21-cm power spectrum at all.

	We showed in Appendix~\ref{app:largescale} that the large-scale 21-cm fluctuations track either the matter or velocity fluctuations,
	\be
	\delta T_{21}(\mathbf x,z) =  b_{\mathcal O}(z) \mathcal O(\mathbf x),
	\ee
	where $\mathcal O=\{\delta_m,\delta_{v}\}$, 
	rescaled with a bias coefficient $b_{\mathcal O}$ that we determined from simulations.
	This required running large-box simulations to extract the bias coefficients, which can be formally written as
	$b_{\mathcal O} (z) = d \overline{T_{21}}(z)/d{\mathcal O}$, and thus represent the response of the 21-cm GS to a larger or smaller matter/velocity fluctuation.
	We can, instead, approximate the result by noticing that a matter overdensity (or velocity decrease) accelerates the evolution of the 21-cm GS. 
	Thus, we can use the time derivative of the global signal as a proxy for the effect of over- or under-densities, effectively writing $b_{\mathcal O}(z)\propto d\overline{T_{21}}/dz$.
	We show the result of this formalism in Fig.~\ref{fig:sigma21approx}, where the cosmic-variance noise from Fig.~\ref{fig:sigma21} is compared against an approximation given by 
	\be
	\sigma_{21,\rm approx} (z) = a \dfrac{d \overline{T_{21}}(z)}{dz},
	\ee 
	with the derivative computed from the GS, and $a=10^{-3}$ is a normalization factor independent of the astrophysical parameters chosen in the simulation, as we have checked it reproduces our calculation of the cosmic variance for the extreme model of Ref.~\cite{Kaurov:2018kez} as well.
	This heuristic approximation provides remarkable agreement with the cosmic variance computed in the main text.

	In addition, we have checked that the correlation between two bins is roughly independent of redshift if expressed in terms of the comoving distance between them, and it can in fact be numerically approximated as $C_{i,j}(\Delta \chi)\approx \exp[-(\Delta \chi/\chi_{\rm corr})^2]$ for $\Delta \chi\leq150$ Mpc, with a correlation length of $\chi_{\rm corr}=60$ Mpc (although of course the shape from Fig.~\ref{fig:Corr21fixz} has the exact result).

	With these tools it is possible to include the cosmic-variance noise in an approximate way for any 21-cm GS model, without knowing the 21-cm fluctuations.
	Given an array of observed frequencies $\nu_i$ (corresponding to $z_i$ and thus with ${\chi}_i$ comoving distances from us), the procedure would be to compute the normalized cosmic covariance matrix as $C_{ij}=\exp[-(\Delta \chi_{ij}/\chi_{\rm corr})^2]$ for $\Delta \chi_{ij} = \chi_i -  \chi_j$, as well as the amplitude of the cosmic-variance noise as 
	$\sigma_{21,i} = a \left. d\overline{T_{21}^{\rm mod}}(z)/dz\right|_{z_i}$, for the input model $\overline{T_{21}^{\rm mod}}$.
	Then that is added to the instrumental noise at each band $\sigma_{{\rm inst},i}$ to find the full covariance matrix as
	\be
	\mathcal C_{ij} = \delta_{ij} \sigma^2_{{\rm inst},i} + \sigma_{21,i} \sigma_{21,j} C_{ij}.
	\ee

	\section{Comparison with simulations}
	\label{app:simulations}

	All the calculations shown in the main text have been analytic, albeit using the 21-cm power spectrum from simulations.
	In this Appendix we confirm our formalism by comparing our results with the variance calculated in 21-cm simulation maps.

	\subsection*{Variance at the same point}
	
	In Fig.~\ref{fig:illustration} we showed how the observed GS varies across our simulation box.
	There we emulated the effect of measuring the GS by averaging our 21-cm simulation over different thin slices.
	In practical terms, we read the full 21-cm map from {\tt 21cmvFAST}, as a square box of size $L_{\rm box}=1.8$ Gpc comoving, where each pixel is $R_{\rm cell}=3$ Mpc comoving, which gives us $N_{\rm slices}=600$ slices in each of the three coordinate directions.
	This 3-Mpc wide band is comparable to the typical band given the resolution of public EDGES data (as for $B=0.4$ MHz the comoving width at $z\sim 20$ is roughly 10 Mpc).
	We will only show simulation results in this Appendix at a single representative redshift $z=16.8$, which lays between the LCE and the EoH, although of course it can be easily generalized to any other $z$.

	We start by writing estimators for the global signal at $z=16.8$ for each of the slices along the line-of-sight direction, labeled by $i$, of the {\tt 21cmvFAST} box by averaging over the 21-cm temperature in the slice,
	\be
	T_{21}^{i,\rm obs} = \dfrac{1}{N_{\rm slices}^2}\sum_{j,k} T_{21}^{i,j,k},
	\ee
	where $j$ and $k$ run over the other two directions, and $T_{21}^{i,j,k} = T_{21}(\mathbf x)$ with $\mathbf x=R_{\rm cell}\times(i,j,k)$.
	Each $T_{21}^{i,\rm obs}$ is, then, an estimate of the global signal for a line-of-sight band of comoving distance $R_{\rm cell}=3$ Mpc.
	This is the quantity we showed in Fig.~\ref{fig:illustration}, averaging over slices in two of the directions of our simulation.
	We can find the true global signal fom the simulation by averaging over all slices
	\be
	\overline{T_{21}} = \dfrac{1}{N_{\rm slices}}\sum_{i} T_{21}^{i,\rm obs} = -91.3 \,\rm mK,
	\ee
	where the last value is for our chosen redshift $z=16.8$.
	From Fig.~\ref{fig:illustration} it is clear that the value that would be measured at different slices can depart from $\overline{T_{21}}$ significantly.
	We find the root mean square of the deviation simply as 
	\be
	\sigma_{21} = \sqrt{N_{\rm slices}^{-1} \sum_i (T_{21}^{i,\rm obs})^2 - \overline{T_{21}}^2 } = 0.53\,\rm mK.
	\label{eq:sigma21direct}
	\ee
	Notice that this variance is a factor of ten larger than our result in the main text, where we estimated the $z=16.8$ the cosmic-variance contribution to the 21-cm GS error budget at $\sigma_{21}\approx 0.05$ mK. 
	That is because the simulation slices are very thin (only 3 Mpc in side) and have different geometry geometry.

	To test our formalism, we will now compute the same variance with our analytic calculation.
	This requires a different window function than used in the main text, 
	as the simulation output has square geometry, as opposed to the spherical shells considered in the main text, which greatly increases the variance in one of the coordinate directions.
	In technical terms, this changes the selection function, which is now 3-Mpc wide along one coordinate direction and 1.8 Gpc on the other two (which breaks isotropy, as in the flat-sky case of Appendix~\ref{app:windows}).
	As before, we start with our key result, Eq.~\eqref{eq:sigma21general}:
	\be
	\sigma_{21}^2 =\int \dfrac{d^3k}{(2\pi)^3} P_{21}(k) \mathcal W_{\rm box}^2(k),
	\ee
	where now
	\be
	\mathcal W_{\rm box}(k) = \mathcal W_{||}(k_{||}) \mathcal W_{\perp}^2(k_\perp),
	\ee
	with 
	\be
	W_{I}(k_{I}) = j_0 \left(\dfrac{k_{I} D_I}{2}\right),
	\ee
	for $I=\{||,\perp\}$, where the two distances are $D_{||}=R_{\rm cell}$ and $D_\perp = L_{\rm box}$.
	This separation of scales allows us to approximately factorize the integral into two parts (since $k_{||}\gg k_\perp$, so $k\approx k_{||}$ for nearly all values of $k$), to find
	\be
	\sigma_{21}^2 \approx I_\perp^2 I_{||},
	\label{eq:varfactorized}
	\ee
	with 
	\be
	I_\perp = \dfrac{1}{\pi} \int dk_\perp [W_{\perp}(k_{\perp})]^2,
	\ee
	and 
	\be
	I_{||} = \dfrac{1}{\pi} \int dk_{||} [W_{{||}}(k_{{||}})]^2 \dfrac{\Delta^2_{21}(k_{||})}{k_{||}^3},
	\ee
	where the integrals only run over positive values of $k_I$.
	This factorization is approximate at the $\sim 10\%$ level, which is enough for our purposes.
	We read the 21-cm power spectrum from the simulation, as before, and evaluate Eq.~\eqref{eq:varfactorized} to find
	\be
	\sigma_{21} = 0.62\,\rm mK,
	\ee
	in good agreement with the value of 0.53 mK obtained directly from simulation slices.

	\begin{figure}[hbtp!]
		\centering
		\includegraphics[width=0.48\textwidth]{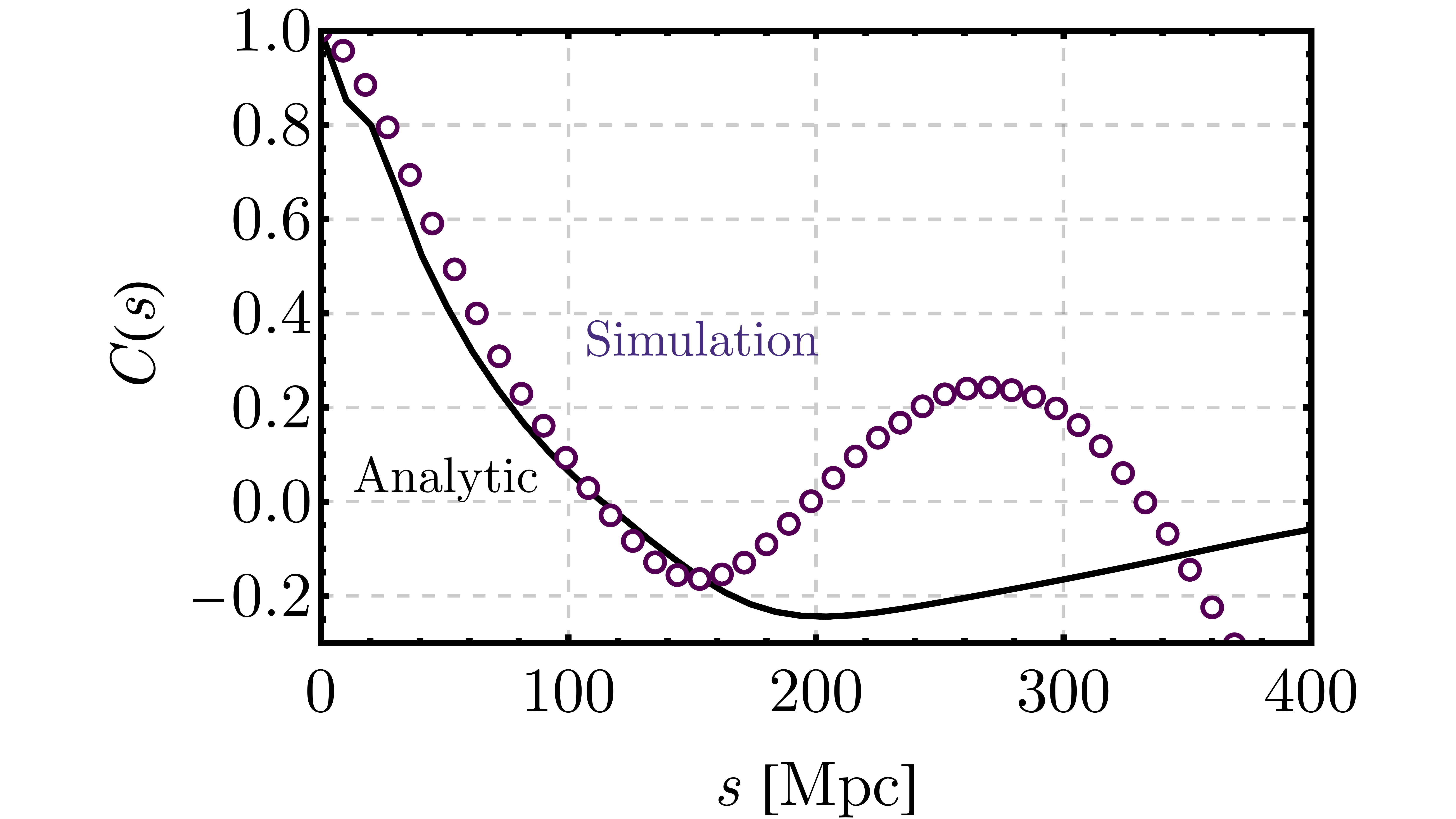}
		\caption{Normalized correlation $C(s)=\sigma_{21}^2(s)/\sigma^2_{21}(0)$ between flat 21-cm slices of our simulated 21-cm map, separated by a distance $s$, at a fixed $z=16.8$.
			Purple circles show the exact correlation in the simulation box, whereas the black curve shows the result of the analytic calculation using the 21-cm power spectrum.
			The disparity between the two curves at large separations $s$ arises due to the periodic initial conditions in the simulation box.
		}	
		\label{fig:Corr21sim}
	\end{figure}

	\subsection*{Correlation between slices}

	Now we move to calculate the covariance of the 21-cm GS, by computing how correlated measurements of the GS are  across slices.
	Since our goal is to test our formalism in the simplest possible way, we will calculate covariances at fixed $z=16.8$, although in reality the power spectrum changes from one $z$ to the next, as described in Appendix~\ref{app:covariance}.
	
	In order to directly compute the correlation between bins in our simulations, we calculate the variance for slices displaced by a comoving distance $s$ as
	\be
	\sigma_{21}^2 (s) = N_{\rm slices}^{-1} \sum_i T_{21}^{i,\rm obs} T_{21}^{i+i_s,\rm obs} - \overline{T_{21}}^2,
	\ee
	where the index $i_s=s/R_{\rm cell}$.
	We show this quantity, normalized to unity at $s=0$ in Fig.~\ref{fig:Corr21sim}.
	This correlation becomes smaller at large $s$, reaching zero at $s\sim 100$ Mpc comoving. 
	For larger separations, however, the correlation appears noisy, and eventually grows above zero again.
	This is a well-known systematic due to the periodic boundary conditions of the simulation boxes, which makes the correlation function depart from the theoretical prediction at large separations $s$~\cite{Sirko:2005uz,Hahn:2011uy} (for instance, the result is the same for $s$ and $L_{\rm box}-s$).
	While this can, in principle, be remedied by sampling the Fourier-space initial conditions differently, we will not attempt to do so here, and instead just focus on the small-$s$ regime.

	The analytic result is obtained from a slightly modified version of Eq.~\eqref{eq:sigma21general},
	\ba
	\sigma_{21}^2(s) &= \VEV{T_{21}(\mathbf x)T_{21}(\mathbf x+\mathbf s)} - \overline{T_{21}}^2 \\\nonumber &= \int \dfrac{d^3k}{(2\pi)^3} P_{21}(k) \mathcal W_{\rm box}^2(k) e^{-i \mathbf k \cdot \mathbf s }.
\end{align}
For a displacement $\mathbf s$ along the line of sight that integral can be approximately factorized as in Eq.~\eqref{eq:varfactorized},
\be
\sigma_{21}^2(s) = I_\perp^2 I_{||}(s),
\ee
with $I_\perp$ is as before, whereas now
\be
I_{||}(s) = \dfrac{1}{\pi} \int dk_{||}  |W_{{||}}(k_{{||}})|^2 \dfrac{\Delta^2_{21}(k_{||})}{k_{||}^3} \cos(k_{||} s).
\ee
We show this quantity (also normalized to $s=0$) in Fig.~\ref{fig:Corr21}.
It vanishes as $s\to \infty$, as expected, and matches well the direct simulation result for $s\lesssim 150$ Mpc.

We emphasize that the calculation performed in this section is different from that outlined in the rest of the text. 
Nevertheless, the close resemblance between our analytic results and those directly computed in the simulation is a good indication that our results are robust.

\end{document}